\documentclass[twocolumn,showpacs,preprintnumbers,amsmath,amssymb]{revtex4}

\usepackage{graphicx}
\usepackage{dcolumn}
\usepackage{bm}

\begin{document}


\title{Negative magnetoresistance and phase slip process in superconducting nanowires}

\author{D.Y. Vodolazov}
\email{vodolazov@ipm.sci-nnov.ru} \affiliation{Institute for
Physics of Microstructures, Russian Academy of Sciences, 603950,
Nizhny Novgorod, GSP-105, Russia}

\date{\today}


\begin{abstract}

We argue that the negative magnetoresistance of superconducting
nanowires, which was observed in recent experiments, can be
explained by the influence of the external magnetic field on the
critical current of the phase slip process. We show that the
suppression of the order parameter in the bulk superconductors
made by an external magnetic field can lead to an enhancement of
both the first $I_{c1}$ and the second $I_{c2}$ critical currents
of the phase slip process in nanowires. Another mechanism of an
enhancement of $I_{c1}$ can come from decreasing the decay length
of the charge imbalance $\lambda_Q$ at weak magnetic fields
because $I_{c1}$ is inversely proportional to $\lambda_Q$. The
enhancement of the first critical current leads to a larger
intrinsic dissipation of the phase slip process. It suppresses the
rate of both the thermo-activated and/or quantum fluctuated phase
slips and results in decreasing the fluctuated resistance.

\end{abstract}

\maketitle

\section{Introduction}

Recently several experimental groups have observed a negative
magnetoresistance (NMR) of the superconducting nanowires at the
temperature lower than the critical temperature
\cite{Xiong,Herzog,Tian1,Rogachev,Arutyunov,Tian2}. In Ref.
\cite{Tian1,Tian2} the authors have directly demonstrated that in
their case the effect is connected with the suppression of the
superconductivity in the bulk superconductors caused by the
applied magnetic field (it was called as 'anti-proximity effect').
They have also found an enhancement of the critical current of the
nanowires when the magnetic field turns bulk superconductors to
the normal state. This result  convinced us that the observed NMR
could be connected with the increase in the critical currents of
the phase slip process. To illustrate how the change of the
critical current value can influence the fluctuated resistance we
use a well-known model of the point-like Josephson junction with a
finite capacitance.

The current-voltage (IV) characteristics of such a junction is
hysteretic and the parameter which governs the IV characteristics
is the ratio between the effective "mass" (which is proportional
to the capacitance C of the junction) and the parameter describing
the effect of intrinsic dissipation (which is inversely
proportional to the resistance R of the junction) \cite{Tinkham}.
In the theory of Josephson junctions this ratio is called the
damping parameter $\beta_c=2eI_cR^2C/\hbar$ ($I_c$ is the critical
current of the Josephson junction with zero capacitance). The
larger $\beta_c$ is the larger is the hysteresis and the smaller
current $I_r$ at which the voltage vanishes in the junction (see
Fig. 1). For small values of $\beta_c$ the current $I_r$ is
practically equal to $I_c$ and the hysteresis is absent.  The
effect of fluctuations leads to the appearance of the finite
resistance at $I<I_r$ and the absence of the hysteresis (phase
slip process does not stop at $I_r<I<I_c$ once it is launched by a
fluctuation) \cite{Tinkham}. If we increase the value of the
current $I_r$ (by {\it decreasing} the resistance of the junction)
the fluctuated resistance at fixed $I\ll I_r$ decreases (compare
gray and black dashed curves in Fig. 1). It is the consequence of
the general rule, that the increase in the intrinsic dissipation W
in the system (in case of Josephson junction $W=V^2/R$) suppresses
both quantum \cite{Schmid2} and thermo-activated fluctuations
\cite{Hanggi}.

\begin{figure}[hbtp]
\includegraphics[width=0.45\textwidth]{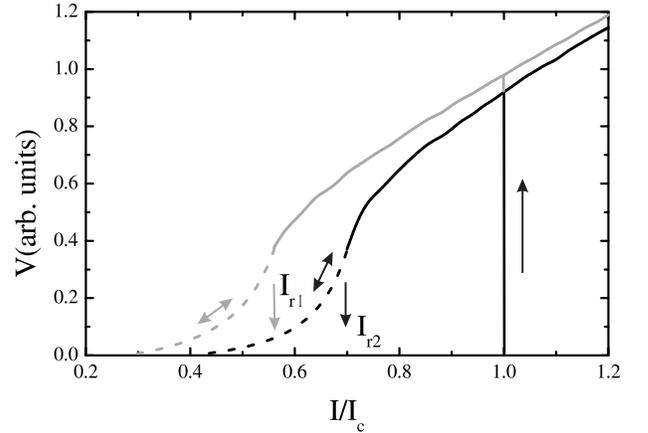}
\caption{Current-voltage characteristics of  Josephson junction
with a finite capacitance and $I_{r1}<I_{r2}$ (parameter
$\beta_c=5$ and 3 respectively) in the case of zero fluctuations
(solid curves). Dashed curves show schematically a non-zero
voltage response at $I<I_r$ due to strong fluctuations. In the
latter case the hysteresis in current-voltage characteristics
disappears.}
\end{figure}

The aim of the present paper is to show that the magnetic field
can enhance the critical currents of the phase slip process and
hence  intensifies the intrinsic dissipation. As in the case of
Josephson junction it leads to the suppression of the rate of
fluctuations and the decrease of the fluctuated resistance.

However, there is a difference between the results in
Refs.\cite{Xiong,Herzog,Rogachev,Arutyunov} and
Refs.\cite{Tian1,Tian2}. In Refs.
\cite{Xiong,Herzog,Rogachev,Arutyunov} the resistance of
superconducting nanowires monotonically decreases with the
increase of H (up to some critical value) while in Refs.
\cite{Tian1,Tian2} the resistance is almost constant at weak
magnetic fields and drops suddenly near the critical field of the
bulk superconductors. Therefore, we conclude that the results are
connected with different mechanisms of the enhancement of the
critical currents in the PS process.

Before considering those mechanisms let us discuss the physical
meaning of the first and the second critical currents of the phase
slip process in superconducting wires.

The paper is organized as follows. In section II we observe the
critical currents of the phase slip process in quasi-1D
superconductors. In section III we study the influence of the
magnetic field on those critical currents. Finally, in section IV
we discuss our results and make a comparison with the experiments
and other theoretical works.

\section{Critical currents of the phase slip process}

It has been known for a long time that the phase slip process in
quasi 1D superconducting wires \cite{self1} is a hysteretic
process (see for example review \cite{Ivlev} and books
\cite{Tidecks,Tinkham}). If we start from the superconducting
state and gradually increase the applied current the
superconducting state becomes unstable at the current $I_{c2}$ (it
is analog of the critical current of Josephson junction $I_c$)
which is equal to the product of the depairing current density
$j_{dep}$ (for wire without defects) on the square of the
cross-section of the sample $S$ . The periodic in time
oscillations of the order parameter in one or several points along
the superconductor destroy the zero-resistance state and bring a
finite resistance (less than normal one) to the system. This
process is called a phase slip (PS) process and the points are
called as  phase slip centers (PSC) \cite{Ivlev,Tidecks,Tinkham}.
If we decrease the current below $I_{c2}$, the phase slip process
can vanish at the current $I_{c1}<I_{c2}$ (which roughly
corresponds to the current $I_r$ of Josephson junction).

The physical origin of the current $I_{c1}$ was clarified in Ref.
\cite{Michotte} using the extended quasi-1D time-dependent
Ginzburg-Landau equations \cite{Kramer,Watts-Tobin}. It has been
found that there are two characteristic times which govern the
dynamic of the order parameter in the phase slip center. These are
the time relaxation of the absolute value of the order parameter
$\tau_{|\psi|}$ and the relaxation time of the phase gradient
$\tau_{\nabla \phi}$ (which is proportional to the momentum of
superconducting electrons).

First characteristic time can be estimated from the time-dependent
Ginzburg-Landau equation for dynamics of $|\Delta|$
\begin{equation}
\tau_{GL}u\sqrt{1+\gamma^2|\Delta|^2}\frac {\partial
|\Delta|}{\partial t}
 =  \xi^2 \frac{\partial^2 |\Delta|}{\partial s^2}+|\Delta|(1-|\Delta|^2-(\xi \nabla \phi)^2),
\end{equation}
where $\Delta=|\Delta|e^{i\phi}$ is the order parameter in
Ginzburg-Landau equations normalized to its equilibrium value at a
specific temperature
$|\Delta|_{eq}=4k_BT_cu^{1/2}(1-T/T_c)^{1/2}/\pi$,
$\tau_{GL}=\hbar/(k_B(T_c-T)u)$ is the Ginzburg-Landau relaxation
time, $\xi=(8k_B(T_c-T)/\pi \hbar D)^{-1/2}$  is a coherence
length (D is a diffusion coefficient),
$\gamma(T)=2|\Delta|_{eq}(T) \tau_E/\hbar$ is the parameter in
time-dependent GL equations, $\tau_E$ is the energy relaxation
time for electrons near Fermi level and $u\simeq 5.79$ is a number
\cite{Kramer,Watts-Tobin}. Numerical analysis shows
\cite{Michotte} that the amplitude of oscillations of $|\Delta|$
in phase slip center is decreasing with the increase of $\gamma$
and it is normally much smaller than $|\Delta|_{eq}$. It allows to
neglect the nonlinear term $|\Delta|^3$ in the right hand side
(RHS) of the Eq. (1) near the core of the phase slip center and it
immediately gives us $\tau_{|\Delta|}\simeq u\gamma\tau_{GL}$. We
can also identify $\tau_{|\Delta|}$ as a relaxation time of the
longitudinal mode in the superconductors $\tau_{|\Delta|} \simeq
\tau_E k_B T_c/\Delta$ \cite{Schmid,Tinkham}.

The second characteristic time $\tau_{\nabla \phi}$ for long wires
$L \gg \lambda_Q$ ($\lambda_Q(\gamma \gg 1) \simeq \sqrt{\gamma/u}
\xi \gg \xi$ is the decay length of the charge imbalance
\cite{Tinkham}) could be estimated by using the Ginzburg-Landau
equation for dynamics of the phase of the order parameter
\cite{Michotte}
\begin{equation}
\frac{\hbar}{2e}\frac {\partial \phi}{\partial t} =
-\varphi+\lambda_Q^2\frac{\partial^2 \varphi}{\partial x^2}.
\end{equation}
For $\lambda_Q \gg \xi $ or $\gamma \gg 1$ the order parameter
{\it mainly} oscillates in a small region (which decreases while
increasing $\gamma$) around the phase slip center with the size
smaller than $\xi$. Therefore, we need to estimate $\tau_{\nabla
\phi}$ in this area. As a result we have $\tau_{\nabla \phi}\sim
\tau_{GL}I_0\xi/I\lambda_Q$ ($I$ is an applied current,
$I_0=S\hbar/2e\tau_{GL}\xi\rho_n$ is proportional to the depairing
current ($I_{c2}=\sqrt(4/27)I_0$), $\rho_n$ is a normal state
resistivity and we take into account that $ -\partial \varphi
/\partial x(x=0)=I_n(x=0)\rho_n/S \sim I\rho_n/S$).

When $\tau_{|\Delta|} \lesssim \tau_{\nabla \phi}$ the phase slip
process is impossible as a periodic one in the time oscillating
process \cite{Michotte} at $I<I_{c2}$. It allows us to estimate
the first critical current of long $L \gg \L_Q$ wires
\begin{equation}
I_{c1} \sim \frac{I_0\tau_{GL}\xi}{\tau_{|\Delta|}\lambda_Q}=
\frac{\hbar}{2e\tau_{\Delta}}\frac{S}{\rho_n\lambda_Q}=
\frac{\hbar}{e\tau_{\Delta}}\frac{1}{R_{PS}},
\end{equation}
where $R_{PS}=2\lambda_Q \rho_n/S$ may be called as a resistance
of the phase slip process \cite{Skocpol}. Note, that in case of
Josephson junction current $I_r$ is also inversely proportional to
the intrinsic resistance \cite{Tinkham}.

Due to the above threshold condition there is a voltage jump
$\Delta V \simeq 1/\tau_{|\Delta|}$ at $I=I_{c1}<I_{c2}$
\cite{Michotte}. If current $I_{c1}$ defined by the above
expression becomes larger than $I_{c2}$ (at $\lambda_Q \lesssim
\xi$ or/and $\tau_{|\Delta|} \lesssim \tau_{GL}$) then the voltage
gradually increases from zero at $I=I_{c2}$ and $\Delta V=0$. In
this limit our estimations for $\tau_{|\Delta|}$ and $\tau_{\nabla
\phi}$ become invalid.

From Eq. (2) it follows that the superconducting electrons with
momentum $p\sim \nabla \phi$ being accelerated by the gradient of
the electrochemical potential $\mu_{e-ch}=e\varphi+\mu_{ch}$ where
$\mu_{ch}$ might be called as a chemical potential of the
superconducting electrons. In Ref. \cite{Schmid} it is shown that
$-\mu_{ch}=e\varphi-\mu_{e-ch}$ is also proportional to the charge
imbalance $Q$ between hole-like and electron-like branches of the
quasiparticle spectrum in superconductors. We may average Eq. (2)
over the period of oscillations of $|\Delta|$ and in the case of
two-dimensional geometry we have
\begin{equation}
\lambda_Q^2\Delta Q-Q=0.
\end{equation}
When the width (w) of the superconducting wire (which is connected
to bulk superconducting reservoirs) is much less than $\lambda_Q$
we can leave only the term with the Laplacian in Eq. (4) near the
ends of the wire and solve 2D Laplace equation. Besides we can
neglect the variation of $Q$ over the width of the nanowire and
solve 1D variant of Eq. (4) in the wire. As a result we obtain the
charge imbalance at the ends of the nanowire $Q_0 \simeq w
Q_c/(\lambda_Q \sinh (L/2\lambda_Q))$($Q_c$ is the charge
imbalance in the phase slip center). Usually $w/L \ll 1$
\cite{Tian1,Tian2} and $Q_0 \ll Q_c$ even for short wires
$\lambda_Q \gg L$. Therefore instead of 2D Eq. (4) we may use (in
the wire) 1D equation
\begin{equation}
\lambda_Q^2\frac{d^2Q}{dx^2}-Q=0,
\end{equation}
with boundary conditions $Q(\pm L/2)=0$, $Q(\pm 0)=\pm Q_c$.

Using Eq. (5) with above boundary conditions it can be found that
the current $I_{c1}$ depends on the length of the nanowire. If the
wire is much longer than the coherence length we may expect that
the order parameter distribution in the core of PSC is not
influenced by the bulk superconductors. Then the dynamics of the
$|\Delta|$ stays the same as for an infinite wire and both
$\tau_{|\Delta|}$ and $\Delta V$ do not suffer any change. From
the solution of Eq. (5) it might be easily seen that the normal
current in the phase slip center (which is proportional to the
applied one) grows with the decrease of $L$ as $I_n(0)\sim
-dQ/dx(x=0)\sim 1/\tanh(L/2\lambda_Q)$ to provide the same charge
imbalance $Q_c\sim \Delta V/2\sim 1/\tau_{|\Delta|}$ near the PSC.
Therefore, the shorter the wire is the larger is $I_{c1}$
($I_{c1}\sim 1/(\lambda_Q\tanh(L/2\lambda_Q))$) and for
sufficiently short wires it becomes equal to $I_{c2}$. We should
note that $I_{c2}$ does not vary while the wire is much longer
than $\xi$. Therefore, we expect that for short wires $L\ll
\lambda_Q$ the hysteresis in current voltage characteristics
disappears \cite{Michotte}.

\section{Effect of the magnetic field}

\subsection{First mechanism}

Let us now discuss how an external magnetic field may influence
$I_{c1}$ and $I_{c2}$. First mechanism comes from the suppression
of the order parameter in the bulk superconductors. In Fig. 2 we
draw the qualitative distribution of $|\Delta|$ and $Q$ at $H=0$,
$H \lesssim H_c^{bulk}$ and $H \gtrsim H_c^{bulk}$. When drawing
the curves we have assumed that the NS boundary forms far from the
ends of the wire at $H \lesssim H_c^{bulk}$ and approaches the
wire at $H\gtrsim H_c^{bulk}$.

Due to conversing  the normal current into the superconducting one
at the NS boundary an additional charge imbalance appears at the
ends of the wire and an effective boundary condition for Eq. (5)
becomes $Q(\pm L/2)=\pm Q_0$. It brings us the following
expression for the first critical current
\begin{equation}
\frac{I_{c1}(H,L)}{I_{c1}(H=0,L=\infty)}=\frac{-Q_0/Q_c+\cosh(L/2\lambda_Q)}{\sinh(L/2\lambda_Q)}
\end{equation}
Current $I_{c1}$ increases with the growth of $H$ because $Q_0$
changes from zero to the maximal value (with the sign opposite to
$Q_c$ - see Fig. 2) when the NS boundary touches the end of the
nanowire (in the latter case the expression for $I_{c1}$ was found
in Ref. \cite{Michotte}). With the increase of a magnetic field
the point where $Q=0$ approaches the center of the wire (see Fig.
2). Hence we can say that the appearance of NS boundaries
effectively shortens the superconductor (in sense that 'space' for
phase slip process decreases) and it is the reason for an
enhancement of $I_{c1}$.

Because the normal current exists on the finite distance from the
NS boundaries inside the superconductor the current $I_{c2}$ is
enhanced too (for wires $L \gg \xi$). Indeed, when the normal
current penetrates far into the sample it decreases the
superconducting component of the current because $I_s+I_n=I$.
Hence, we need a larger applied current $I$ to satisfy the
condition $I_{c2}=I_s$.
\begin{figure}[hbtp]
\includegraphics[width=0.45\textwidth]{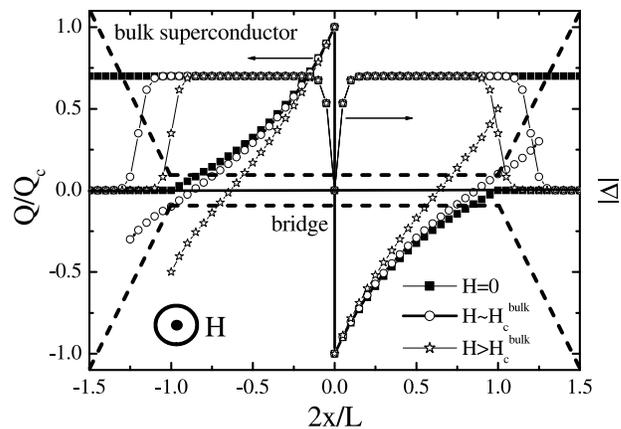}
\caption{Schematic distribution of the charge imbalance and order
parameter in the superconducting bridge ($L=4\lambda_Q$) with one
phase slip center at different values of the magnetic field.}
\end{figure}

The characteristic length of the discussed mechanism is the decay
length of the charge imbalance. It means that the effect exists
only in relatively short wires $L\lesssim \lambda_Q$. Besides the
nanowire should not be wide, otherwise the critical field of the
wire $H_c\sim 1/(\xi w)$ and a bulk superconductor $H_c^{bulk}\sim
1/(\xi^2)$ become close to each other and the magnetic field
strongly suppresses $|\Delta|$  in the wire (in framework of GL
model $|\Delta|=(1-(H/H_c)^2)^{1/2}$). It leads to increasing
$\tau_{|\Delta|}\sim 1/(1-(H/H_c)^2$ and $\lambda_Q \sim
1/|\Delta|^{1/2}$ \cite{Michotte} and hence to decreasing $I_{c1}$
if the effect of the NS boundaries is weak. In derivation of Eq.
(6) we suppose that both $\tau_{|\Delta|}$ and $\lambda_Q$ do not
depend on the magnetic field. It is true if $H\ll H_c$ and at some
additional conditions (see the subsection below).

\subsection{Second mechanism}

The second mechanism of a variation of $I_{c1}$ comes from the
dependence of pair-breaking mechanisms on the magnetic field due
to an orbital effect \cite{Schmid,Schon}. In Ref. \cite{Schmid}
the decreasing $\lambda_Q$ with the increase of the applied
magnetic field was predicted for weak magnetic fields. Because the
first critical current depends on $\lambda_Q$ as $I_{c1} \sim
1/\lambda_Q$ we can expect that $I_{c1}$ increases in weak
magnetic fields. Note that in contrast to the first mechanism
$I_{c2}$ decreases in this case.

The quantitative expression for $\lambda_Q(H)$ was found in Ref.
\cite{Schmid}
\begin{equation}
\lambda_Q(H)=\frac{4Dk_BT\tau_E}{\pi |\Delta|(H)}
\left(\frac{1}{1+\gamma(0)(H/H_c(0))^2}+\frac{1}{\gamma(T)^2}
\right)
\end{equation}
for temperatures close to $T_c$. The physical reason for the
dependence of $\lambda_Q$ on H is the following. The decay of the
charge imbalance in superconductors occurs due to Andreev
reflection process (for quasiparticles with energy less than
$|\Delta|$ near the NS boundary) or/and due to an inelastic
electron-phonon interaction (for quasiparticles with the energy
larger than $|\Delta|$ and along the whole superconductor). A weak
magnetic field almost does not influence the order parameter, but
it smears the density of states of the quasiparticles
\cite{Anthore} and makes possible Andreev reflection process of
the quasiparticles with the energy larger than $|\Delta|$. It
provides a faster relaxation of the charge imbalance and decreases
$\lambda_Q$. A high magnetic field strongly suppresses the order
parameter and makes the contribution to Andreev reflections
smaller. It increases effective $\lambda_Q$.

We study this effect in the framework of the time-dependent GL
equations. We use the field and temperature dependent parameter
$\gamma_H=\gamma/(1+\gamma(H/H_c)^2(1-T/T_c)^{1/2})^{1/2}$ in Eq.
(1,2) and add the term ($-|\Delta|(H/H_c)^2$) in the right hand
side of Eq. (1) \cite{Michotte} (we use dependence
$H_c(T)=H_c(0)\sqrt{1-T/T_c}$). For example for tin ($\gamma\simeq
100(1-T/T_c)^{1/2}$) the effect is expected to be weak and for
aluminium ($\gamma\simeq 10^4(1-T/T_c)^{1/2}$) it should be
noticeable already at the temperature close to $T_c$.
\begin{figure}[hbtp]
\includegraphics[width=0.45\textwidth]{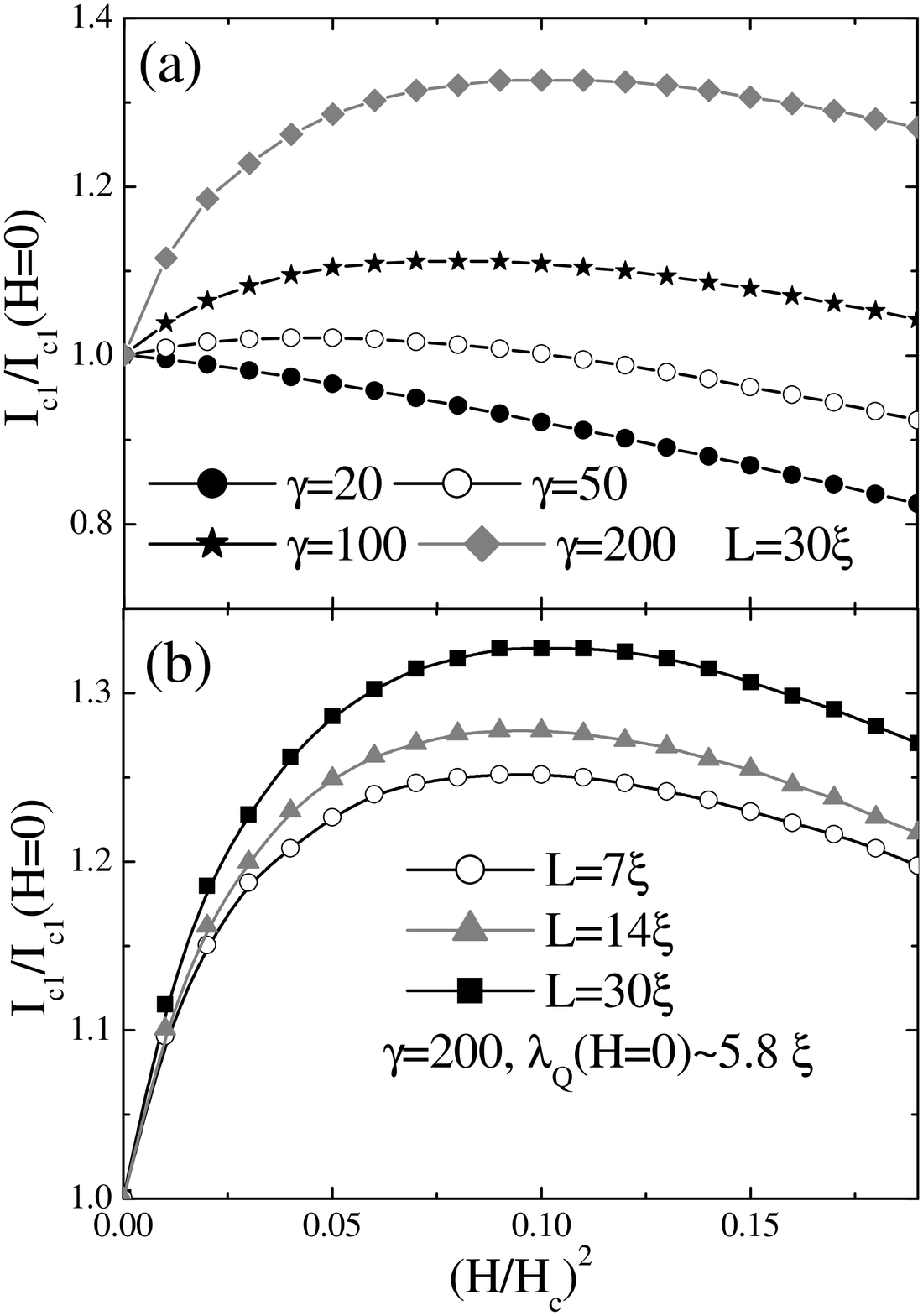}
\caption{(a)Dependence of $I_{c1}$ on the magnetic field which was
found from the numerical solution of Eqs. (1,2) with a field
dependent parameter
$\gamma_H=\gamma/(1+\gamma(H/H_c)^2(1-T/T_c)^{1/2})^{1/2}$ at
different values of $\gamma$. (b) Dependence of $I_{c1}$ on the
magnetic field at the fixed value of $\gamma=200$ and different
lengths of the superconducting wire.}
\end{figure}

In our calculations we take $T=0.9T_c$ and several values of the
parameter $\gamma$ (see Fig. 3(a)). At a weak magnetic field
$I_{c1}$ increases if the parameter $\gamma$ is sufficiently high.
It occurs not only due to decreasing $\lambda_Q \sim
\sqrt{\gamma_H}$ but also due to decreasing $\tau_{|\Delta|}\sim
\gamma_H$. The large magnetic field suppresses the order parameter
considerably and it leads to decreasing $I_{c1}$ and $I_{c2}$ (see
the end of Sec. III A). Because for short wires $I_{c1}$ does not
depend on $\lambda_Q$ the enhancement of the $I_{c1}$ due to the
discussed mechanism should be weakened in such a samples. The
numerical calculations confirmed that statement (see Fig. 3(b)).

\section{Discussion}

We can present simple interpretations of our results. In long ($L
\gg L_Q$) wires the resistance of the phase slip process $R_{PS}$
is proportional to $2\lambda_Q\rho_n/S$ \cite{Skocpol}. In short
($L \lesssim L_Q$) wires $R_{PS}$ is proportional to the length of
the wire $R_{PS}\simeq L\rho_n/S$ because the electric field is
close to zero in bulk superconductors. When a short wire is
bounded by the normal metal the resistance of the phase slip
process {\it decreases} by the value which is proportional to the
resistance of the NS boundaries $R_{PS}\simeq L\rho_n/S-2R_{NS}$.
When one applies a magnetic field  $R_{PS}$ can decrease due to
decreasing $\lambda_Q$ (in long wires) or due to the appearance of
$R_{NS}$ (in short wires) and as in the case of Josephson junction
current $I_{c1}\sim \Delta V/R_{PS}$ increases. The intrinsic
dissipation $W \sim V^2 /R_{PS}$ increases too and it suppresses
the fluctuations of the phase of the order parameter and results
in decreasing the fluctuated resistance.

The extended time-dependent Ginzburg-Landau equations are valid
for a narrow temperature interval close to the critical
temperature $|T_c-T|<\hbar /k_B \tau_E$ where
$\xi(T)>L_E=\sqrt{D\tau_E}$. However, we may suppose that traces
of the high temperature dynamics should exist at low temperatures
too. It was experimentally found for a tin nanowire with $L=6\mu
m$ the presence of several phase slip centers (see Fig. 3b in Ref.
\cite{Tian3}) at $T=0.5K\simeq 0.12 T_c$. First phase slip brings
a finite resistance which is equivalent to the resistance of the
piece of the wire with the length about 1.5 $\mu m$ and hence
using SBT theory \cite{Skocpol} we obtain $\lambda_Q(Sn)\simeq 750
nm$. It is much larger than the coherence length in tin at this
temperature ($\xi(Sn,T=0) \lesssim 55 nm$).

In Ref. \cite{Vodolazov} the S-behavior of the current voltage
characteristics of superconducting nanowires was found in a
voltage driven regime at low temperatures. Time-dependent
Ginzburg-Landau equations with a large value of $\gamma$ give
qualitatively the same result \cite{Michotte,Vodolazov}. The
voltage jump $\Delta V$ was extracted from the experimental data
and showed a qualitative agreement with the theoretical
temperature dependence $\Delta V(T)$ \cite{Michotte}.

In the recent paper \cite{Menten} the pronounced hysteresis of the
IV characteristic of the Pb nanowire with several phase slip
centers was observed at a low temperature. In the preceding
experiments \cite{Vodolazov,Michotte} the hysteresis was hidden by
a strong external noise. Proper filtering suppressed the noise
\cite{Menten} and revealed the hysteretic behavior with well
identified currents $I_{c1}$ and $I_{c2}$.

The above experiments support the idea that the low-temperature
properties of the phase slip process resemble such ones at high
temperatures. Therefore, we expect that our results based on the
numerical solution of Eqs. (1,2) and Eqs. (4,5) are applicable for
{\it qualitative} analysis of the phase slip process at low
temperatures.

The magnetic field dependence of the first and second mechanisms
of the current enhancement are rather different. In the case of
the first mechanism the enhancement occurs at the moment when bulk
superconductors switch to the normal state. In the second
mechanism the current $I_{c1}$ increases smoothly with H (see Fig.
3). Therefore, we believe that the first mechanism may be
responsible for the effects observed in Refs.\cite{Tian1,Tian2}
while the second mechanism is connected with the experiments in
Refs. \cite{Xiong,Herzog,Rogachev,Arutyunov}.

Indeed, the theory for the first mechanism clarifies why the
'anti-proximity effect' \cite{Tian1,Tian2} is weakened in long Zn
nanowires with $L=30 \mu m$ ($\lambda_Q(Zn) \sim 22 \mu m$) and is
absent in Sn nanowires with lengths $L=6-30 \mu m$ ($\lambda_Q(Sn)
\sim 750 nm$ - see our estimation above). They seem to be too long
to observe the 'anti-proximity effect'.

In Ref. \cite{Tian2} direct measurements demonstrated an increase
of the critical current in the superconducting nanowires. We
identify this current as the first critical current of the phase
slip process. The second critical current was not observed in the
experiment due to a high rate of fluctuated PSC which reveals
itself in the finite resistance of the wire at $I \to 0$. The
absence of hysteresis is also typical for Josephson junction with
a high rate of fluctuations \cite{Tinkham}.

To observe the first mechanism of NMR (due to current enhancement)
the width w or diameter d of the wire should be relatively small
to provide a condition $H_c\sim 1/(w,d) \gg H_c^{bulk}$. Otherwise
the magnetic field $H=H_c^{bulk}$ strongly suppresses the order
parameter in the wire. It leads to decreasing both critical
currents $I_{c1} $, $I_{c2}$ (see and of Sec. III A) and results
in a negligible effect of the normal boundaries. We believe that
it is the reason for the observed dependence of the
'anti-proximity effect' on the width of the nanowires in Ref.
\cite{Tian1}. The authors found no effect for a wire with $d=70 nm
\sim \xi/2 $ and pronounced effect for a nanowire with $d=40\sim
\xi/4$.
\begin{figure}[hbtp]
\includegraphics[width=0.45\textwidth]{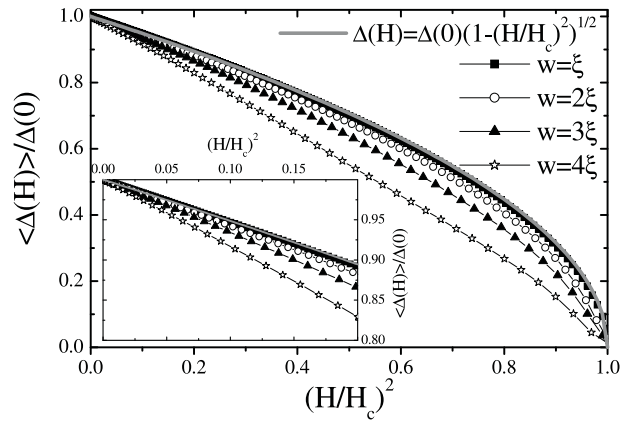}
\caption{ Dependence of the order parameter averaged over the
width of the film on the applied magnetic field for different
values of $w$ (the result is obtained in framework of
Ginzburg-Landau equations). It shows a relatively stronger
suppression of $|\Delta|$ in films with $w>2\xi$ than in narrow
films $w<2\xi$ for the same ratio $H/H_c$.}
\end{figure}

The maximal enhancement of the current $I_{c1}$ does not depend on
$w,d$ (while $w,d \lesssim \xi$) in the second mechanism. The
invariance is connected with a universal dependence of the order
parameter on a magnetic field
$\Delta(H)=\Delta(0)(1-(H/H_c)^2)^{1/2}$ for narrow samples. For
wider samples $w>\xi$ the order parameter decreases faster with
the growing ratio of $H/H_c$ (see Fig. 4) and the enhancement of
$I_{c1}$ becomes weaker. For very wide samples the vortices enter
the sample at $H \ll H_c\sim H_{c2}$ and that suppresses the order
parameter even stronger. The conclusion is that the second
mechanism of enhancement of $I_{c1}$ is maximal in narrow quasi-1D
samples with $w,d \lesssim \xi$.

We can speculate that the observed in Ref. \cite{Xiong}
suppression of the NMR for wide bridges is connected with
exceeding width of the sample over the coherence length. For
example the minimal NMR occurs for a sample with $w=100 nm$ (see
Fig. 4 of Ref. \cite{Xiong}). Taking into account the coherence
length of bulk lead ($\xi(0)=83 nm$) and a dirty limit expression
for the coherence length $\xi\sim\sqrt{\hbar D/|\Delta|}$ we have
$\xi(T=1.6 K) \simeq 50 nm$ (for samples with $T_c\simeq 2.5 K$).
Therefore, for the widest sample our rough estimation gives
$w\simeq 2\xi$. Besides from our theory of the second mechanism of
NMR it follows that the maximal current enhancement (or maximal
suppression of the resistance) occurs at $H^*\sim H_c\sim 1/(w,d)$
for $w,d \lesssim \xi$ (see Fig. 3a). In Refs.
\cite{Xiong,Rogachev} qualitatively the same dependencies were
observed.

Our both mechanisms give a suppression of NMR at approaching $T_c$
(similar to experiments
\cite{Xiong,Herzog,Tian1,Rogachev,Arutyunov,Tian2}). The first
mechanism becomes noticeable if the length of the wire is not very
short. Otherwise the current $I_{c1}$ is equal to $I_{c2}$ and
$I_{c2}$ decreases for wires with $L \lesssim 4\xi(T)$ due to a
strong proximity effect from the normal banks.

The second mechanism of NMR is effective only when the term
$\gamma(0) (H/H_c(0))^2$ is larger than the unity (see Eq. (7)).
It is obvious that for some temperatures close to $T_c$ the above
condition fails because $H<H_c(T)\ll H_c(0)$. Besides with
increasing temperature the charge imbalance length increases as
$\lambda_Q \sim \lambda_Q(0)(1-T/T_c)^{-1/2}$ \cite{Tinkham} and
the length of the sample becomes smaller than $\lambda_Q$. It also
suppresses the second mechanism of NMR (see Fig. 3b).

The proposed mechanisms of the negative magnetoresistance are
different from those studied in Refs. \cite{Fu,Pesin}. The authors
of work \cite{Fu} supposed that {\it additional} resistance due to
NS boundaries stabilizes the superconducting phase. In our
approach the intrinsic dissipation grows due to {\it decreased}
intrinsic resistance of the phase slip process and it suppresses
the fluctuations in the system.

In Ref. \cite{Pesin} a new channel of dissipation in
superconducting wires was proposed which can be suppressed by an
external magnetic field. It would be interesting to compare the
contributions of that channel and the second mechanism of NMR
studied in the present paper.

Our mechanisms of current enhancement are also rather different
from a critical current enhancement predicted in Refs.
\cite{Kharitonov,Wei,Kharitonov2}. In those works the enhancement
of current $I_{c2}$ (using our terminology) was found due to {\it
suppression} by applied magnetic field the pair-breaking resulting
from total 'spin-flip'+'non-spin-flip' rate \cite{Kharitonov2}.
This process leads to decreasing the current $I_{c1}$ because
$\lambda_Q$ grows according to that effect. In our second
mechanism the increased pair-breaking (due to orbital effect) {\it
decreases} current $I_{c2}$ but {\it enhances} the current
$I_{c1}$.

Our first mechanism has rather different behavior on magnetic
field and cannot be confused with theory of
Refs.\cite{Wei,Kharitonov2}. The direct way to distinguish among
the second mechanism of a current enhancement is predicted in our
paper and Refs. \cite{Wei,Kharitonov2} is to study the samples of
different lengths and widths (with other close parameters). In
contrast to Refs. \cite{Wei,Kharitonov2} the current enhancement
in our second mechanism depends on the length of a nanowire and
becomes weaker for wires with the length $L<2\lambda_Q$. The
second difference is that the maximal current enhancement does not
depend on the width of a wire for narrow samples $w \lesssim \xi$
while in \cite{Wei,Kharitonov} the strong dependence on $w$ was
predicted even for such a narrow wires.

\begin{acknowledgements}

The work was supported by the INTAS (Grant N 04-83-3139) and the
program RAS 'Quantum Macrophysics'. Author thanks to A. S.
Mel'nikov for useful discussions.

\end{acknowledgements}


\begin{references}

\bibitem{Xiong} P. Xiong, A. V. Herzog and R. C. Dynes,
Phys. Rev. Lett. {\bf 78}, 927 (1997).

\bibitem{Herzog} A. V. Herzog, P. Xiong,, and R. C. Dynes,
Phys. Rev. B {\bf 58}, 14199 (1998).

\bibitem{Tian1} M. L. Tian, N. Kumar, S. Y. Xu, J. G. Wang, J. S. Kurtz, and M.
H. W. Chan, Phys. Rev. Lett. {\bf 95}, 076802 (2005).

\bibitem{Rogachev} A. Rogachev, T.-C. Wei, D. Pekker, A. T. Bollinger, P. M.
Goldbart, and A. Bezryadin, Phys. Rev. Lett. {\bf 97}, 137001
(2006).

\bibitem{Arutyunov} K. Arutyunov (private communication).

\bibitem{Tian2} M. Tian, N. Kumar, J. G. Wang, S. Y. Xu and M. H. W.
Chan, Phys. Rev. B {\bf 74}, 014515 (2006).

\bibitem{Tinkham} M. Tinkham, {\it Introduction to
superconductivity}, (McGraw-Hill, NY, 1996).

\bibitem{Schmid2} A. Schmid, Phys. Rev. Lett. {\bf 51}, 1506 (1983).

\bibitem{Hanggi} P. H\"{a}nggi, P. Talkner, and M. Borkovec, Rev. Mod. Phys. {\bf
62} 251 (1990).

\bibitem{self1} Quasi 1D wire is the three-dimensional sample of finite
length and finite cross-section with width (or diameter) less than
the coherence length $w,d \lesssim \xi$. Last condition allows us
to neglect the variations of the order parameter over the
transverse(radial) coordinates and consider the dependence only on
the longitudinal (along the wire) coordinate.

\bibitem{Ivlev} B.I. Ivlev and N.B. Kopnin, Adv. Phys. {\bf 33}, 80 (1984).

\bibitem{Tidecks} R. Tidecks, {\it Current-induced nonequilibrium
phenomena in quasi-one-dimensional superconductors}, (Springer,
Berlin, 1990).

\bibitem{Michotte} S. Michotte, S. Matefi-Tempfli, L. Piraux, D.
Y. Vodolazov, and F. M. Peeters, Phys. Rev. B {\bf 69} 094512
(2004).

\bibitem{Kramer} L. Kramer and R.J. Watts-Tobin, Phys. Rev. Lett.
{\bf 40}, 1041 (1978).

\bibitem{Watts-Tobin} R.J. Watts-Tobin, Y. Kr\"ahenb\"uhl, and L. Kramer,
 J. Low Temp. Phys. {\bf 42}, 459 (1981).

\bibitem{Schmid} A. Schmid and G. Sch\"on, J. Low Temp. Phys. {\bf
20}, 207 (1975).

\bibitem{Skocpol} W. J. Skocpol, M. R. Beasley, and M. Tinkham,
J. Low Temp. Phys. {\bf 16}, 145 (1974).

\bibitem{Schon} G. Sch\"{o}n and V. Ambegaokar, Phys. Rev. B {\bf
19}, 3515 (1979).

\bibitem{Anthore} A. Anthore, H. Pothier,  and D. Esteve, Phys. Rev. Lett. {\bf 90},
127001 (2003).

\bibitem{Tian3} M. L. Tian, J. G. Wang, J. S. Kurtz, Y. Liu, and M. H. W. Chan,
Phys. Rev. B {\bf 71}, 104521 (2005).

\bibitem{Vodolazov} D. Y. Vodolazov, F. M. Peeters, L. Piraux, S. Matefi-Tempfli, and S. Michotte,
Phys. Rev. Lett. {\bf 91} 157001 (2003).

\bibitem{Menten} F. de Menten de Horne, L. Piraux, and S. Michotte,
Physica C (Proceedings of M2S-HTSC-VIII conference, Dresden,
Germany, 9-14 July, 2006).

\bibitem{Fu} H. C. Fu, A. Seidel, J. Clarke, and D.-H. Lee, Phys.
Rev. Lett. {\bf 96}, 157005 (2006).

\bibitem{Pesin} D. A. Pesin and A. V. Andreev, Phys. Rev. Lett. {\bf
97}, 117001 (2006).

\bibitem{Octavio} M. Octavio, W. J. Skocpol, and M. Tinkham, Phys. Rev. B{\bf 17}, 159
(1978).

\bibitem{Kharitonov} M. Yu. Kharitonov and M. V. Feigel'man, JETP Lett. {\bf 82}, 421
(2005); cond-mat/0504433.

\bibitem{Wei} T. C. Wei, D. Pekker, A. Rogachev,
A. Bezryadin, and P. M. Goldbart, Europhys. Lett. {\bf 75}, 943
(2006).

\bibitem{Kharitonov2}  M. Yu. Kharitonov and M. V. Feigel'man, cond-mat/0612455.

\end{references}
\end{document}